\documentclass[type=submission,name=JSS]{oae}

\usepackage{siunitx}
\usepackage{booktabs}
\usepackage{caption}
\usepackage{subcaption}
\usepackage{hyperref}
\usepackage{newtxmath}
\usepackage{braket}

\documentsetup{
  title = {Quantum-Secured Data Centre Interconnect in a field environment},
  authors = {
    author1 = {
      name         = {Kaiwei Qiu},
      affiliations = {affil1},
    },
    author2 = {
      name         = {Jing Yan Haw},
      affiliations = {affil2},
    },
    author3 = {
      name         = {Hao Qin},
      affiliations = {affil2},
    },
    author4 = {
      name         = {Nelly H. Y. Ng},
      affiliations = {affil1},
    },
     author5 = {
      name         = {Michael Kasper},
      affiliations = {affil5},
    },
     author6 = {
      name         = {Alexander Ling},
      affiliations = {affil2,affil3},
    },
  },
  affiliations = {
    affil1 = {
      School of Physical and Mathematical Sciences, Nanyang Technological University, 21 Nanyang Link, Singapore 637371.
    },
    affil2 = {
     Centre for Quantum Technologies, National University of Singapore, 3 Science Drive 2, Singapore 117543.
    },
    affil3 = {
     Department of Physics, National University of Singapore, 2 Science Drive 3, Singapore 117551.
    },
    affil4 = {
      Department of Electrical and Computer Engineering, National University of Singapore, Singapore 117583.
    },
    affil5 = {
      Fraunhofer Singapore Research Centre@NTU, Nanyang Technological University, 62 Nanyang Drive, Singapore 637459.
    },
  },
  correspondence-to = {
    Dr. Jing Yan Haw, Centre for Quantum Technologies, National University of Singapore, 3 Science Drive 2, 117543 Singapore. E-mail: jy.haw@nus.edu.sg; ORCID: 0000-0002-8318-9245; 
    Dr. Hao Qin, Centre for Quantum Technologies, National University of Singapore, 3 Science Drive 2, 117543 Singapore. E-mail: hao.qin@nus.edu.sg; 
  },
  short-author      = {\textit{Qiu et al}.},
  journal           = {Journal of Surveillance, Security and Safety},
  short-journal     = {J Surveill Secur Saf},
  how-to-cite       = {},
  received          = {}, 
  first-decision    = {},
  revised           = {},
  accepted          = {},
  published         = {},
  academic-editor   = {},
  copy-editor       = {},
  production-editor = {},
  doi               = {},
  year              = {},
  volume            = {},
  end-page          = {},
}

\begin{document}
\maketitle
\begin{abstract}

In the evolving landscape of quantum technology, the increasing prominence of quantum computing poses a significant threat to the security of conventional public key infrastructure. Quantum key distribution (QKD), an established quantum technology at a high readiness level, emerges as a viable solution with commercial adoption potential. QKD facilitates the establishment of secure symmetric random bit strings between two geographically separated, trustworthy entities, safeguarding communications from potential eavesdropping. In particular, data centre interconnects can leverage the potential of QKD devices to ensure the secure transmission of critical and sensitive information in preserving the confidentiality, security, and integrity of their stored data. In this article, we present the successful implementation of a QKD field trial within a commercial data centre environment that utilises the existing fibre network infrastructure. The achieved average secret key rate (SKR) of \SI{2.392}{kbps} and an average quantum bit error rate (QBER) of less than \SI{2}{\%} demonstrate the commercial feasibility of QKD in real-world scenarios. As a use case study, we demonstrate the secure transfer of files between two data centres through the Quantum-Secured Virtual Private Network, utilising secret keys generated by the QKD devices.

\end{abstract}

\section{1. Introduction}
As the quantum technology landscape evolves, there is recognition of a threat on the horizon: quantum computing poses a threat to the security of existing asymmetric encryption techniques \cite{mavroeidis2018impact,chen2016report}. In order to circumvent possible breaches of long-term information security, it would be prudent to begin evaluating possible quantum-safe technologies as candidate solutions. An existing quantum technology that is of relevant high readiness level and ready for commercial adoption is quantum key distribution (QKD). QKD enables two distant, honest parties to work together to create shared symmetric random bit strings that remain secure from a potential eavesdropper. Current-day key establishment protocols, such as the RSA (Rivest Shamir Adleman), rely on the assumption that an adversary has limited computational power relative to the hardness of a mathematical problem. In contrast, the security of QKD protocol can be proven even against an eavesdropper with unbounded computational power (including quantum computers). This security is called information-theoretic security (ITS)\cite{PhysRevA.72.012332}. Furthermore, QKD is among the first technologies based on quantum information that is commercially available and has been deployed in fibre networks and free space setups worldwide. 

The overall performance of QKD hardware in a commercial environment is indicated by two key parameters: the secret key rate (SKR) and quantum bit error rate (QBER) \cite{Stucki_2011}.
The SKR indicates the achieved rate of secret keys produced by the QKD devices, whereas QBER provides an error percentage for quantum signal transmissions between the QKD end nodes over a quantum channel. (For a continuous variable type QKD system, the quantum channel is characterised by the channel transmission and the excess noise~\cite{peev2009secoqc}.) To get a glimpse of the performance of production-grade QKD deployments in a commercial environment, as early as 2008, one of the QKD networks in the SECOQC project recorded an average SKR of \SI{3.1}{kbps} and QBER of \SI{2.6}{\%} over a \SI{33}{km} fibre distance with a fibre loss of \SI{7.5}{dB} in Vienna \cite{peev2009secoqc}. In Switzerland, an average SKR of \SI{2.5}{bps} and QBER of \SI{5}{\%} was achieved with the QKD system that uses the coherent one-way (COW) protocol \cite{gisin2004towards, COWprotocol1} over a fibre link of \SI{150}{km} with loss of \SI{43}{dB} from Neuchatel to Geneva \cite{stucki2009continuous}. In 2012, a long-term field demonstration of a QKD network that links two metropolitan cities with a trusted node was established in China, where the longest link spans a distance of \SI{85.1}{km}, with a fiber loss of \SI{18.4}{dB}, recorded an average SKR of \SI{0.77}{kbps} and QBER of \SI{5.26}{\%} for decoy state BB84 protocol \cite{Wang:14}. More recently, the quantum network established in Cambridge, United Kingdom, which uses BB84 protocol \cite{bennett2014quantum}, recorded an average SKR of \SI{2580}{kbps} and an average QBER below \SI{2.5}{\%} over a fibre distance of \SI{10.6}{km} with a loss of \SI{3.9}{dB} \cite{dynes2019cambridge}. Beyond the general performance of the QKD system, the potential for seamless integration between QKD technology and encryption-based applications offers a captivating prospect for its potential commercial use cases. For instance, the generated QKD keys can be used to establish point-to-point quantum-secure communication links to transfer data \cite{Stucki_2011,Braun:21} and perform video conferencing \cite{Sasaki:11}. Furthermore, the possibility of a secure financial transaction over a quantum-secure optical channel with a QKD system using BB84 protocol has been demonstrated by a financial bank in a lab environment recently \cite{Pistoia_2023}.

Securing the transmission of private and sensitive data is an important application for the integration of QKD devices, especially for critical information infrastructure. Data centres, in particular, can leverage the QKD devices for this very purpose \cite{cao2022evolution,liu2022towards,jain2023quantum}. They are infrastructures for companies or organisations to house their IT equipment, allowing them to perform tasks such as data storage, remote applications, accessing cloud computation services. The data traffic experienced by these data centres is growing rapidly and a forecast done by Cisco in 2018 shows that this traffic will reach 19.5 zettabytes by the year 2021 globally \cite{Cisconews-2018}. This is expected to increase even more in recent years given the increasing demand for cloud storage and advancement of cloud services, and more data centres will be needed to handle these large amounts of data traffic. In addition, communication between these data centres is required to fetch and retrieve data seamlessly. This communication link is called the Data Centre Interconnect and it is responsible for establishing a network connecting all data centres together \cite{Cheng:18}. Most often, these interconnects are established using a Virtual Private Network (VPN), and it is crucial for them to be secure to prevent compromising the confidentiality, security and integrity of the data within the data centres \cite{jain2023quantum}. Therefore, various stakeholders have begun to employ quantum-safe cryptography solutions by deploying QKD devices to provide secure keys for encrypting these interconnects for secure data transfer \cite{zatoukal2021openqkd,cao2022evolution} and cloud computing \cite{AWScloud}. In addition, an essential consideration when integrating QKD devices into the data centre is that this process should not require any major modification to their current network configuration, or building a new fibre infrastructure specifically for the QKD devices; in fact, these QKD devices should be seen as an upgrade to their existing interconnect network for quantum-safe readiness. Field trials done under this consideration will provide a good indication of the commercial viability of QKD devices.

In this article, we report the demonstration of a successful QKD field trial in a commercial data centre environment over existing fibre network infrastructure. This deployment was conducted by the National Quantum-Safe Network of Singapore (NQSN), a nationwide testbed for quantum-safe technology, in collaboration with Singapore Technologies Telemedia Global Data Centres (STT-GDC). The goal was to examine the technical feasibility and reliability of production-grade QKD equipment in the context of Singapore's commercial operating environment. We covered the entire deployment life cycle: starting from the installation of QKD devices in the data centres and network equipment setup, to the subsequent monitoring of the physical layer of the quantum network. At the same time, as a case study, we explore the possibility of utilising the keys generated via the QKD devices to create a quantum-secured virtual private network to demonstrate secure data transmission between two interconnected data centres. 

\section{2. Methods}
\subsection{2.1 QKD Background and QKD Protocol}

\begin{figure}[!htbp]
    \centering
    \includegraphics[width=0.9\textwidth]{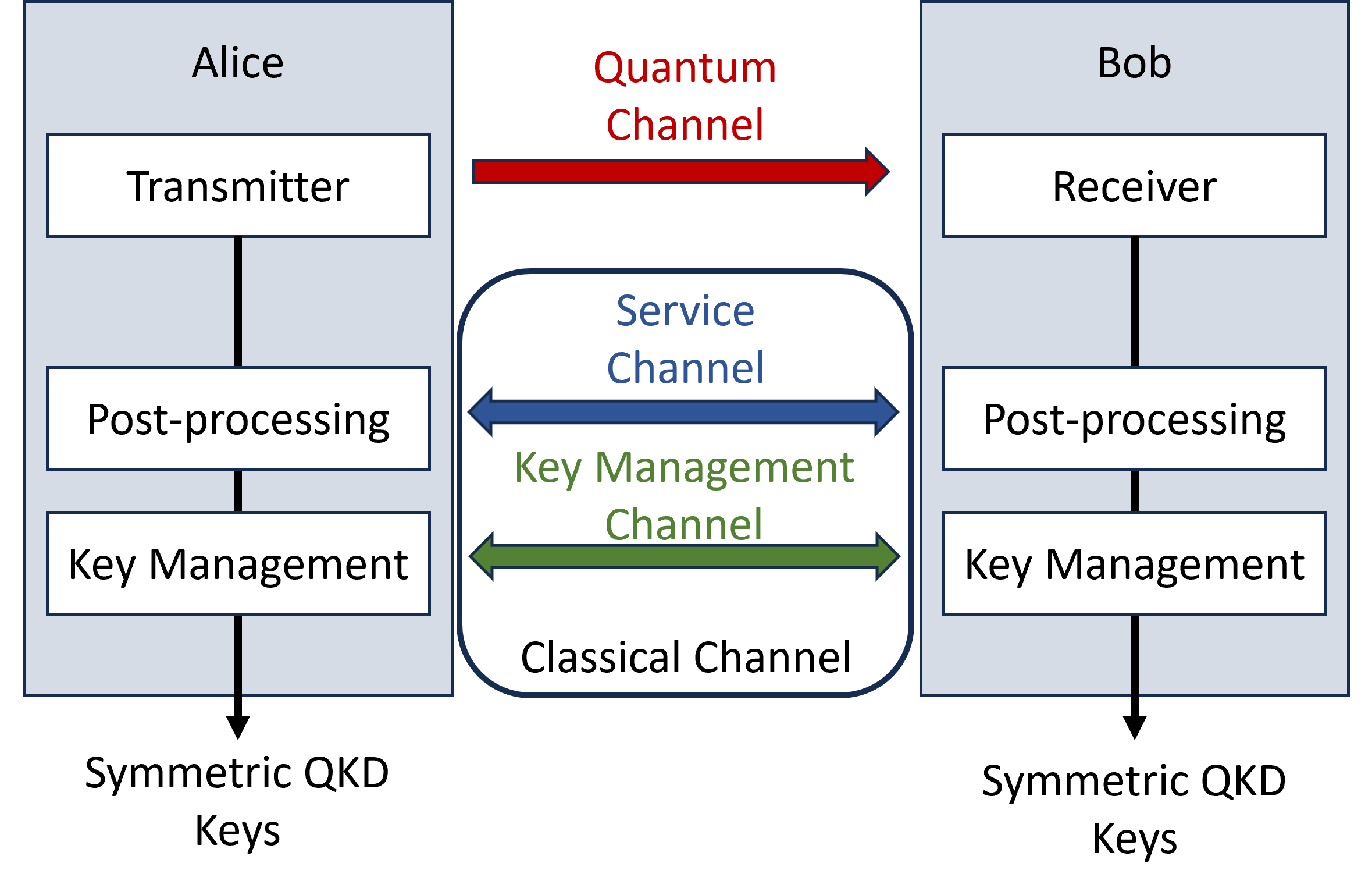}
    \caption{The basic components for a QKD system. The quantum channel (red arrow) exchanges the raw keys between Alice and Bob. Afterwards, post-processing takes place over the classical service channel (blue arrow), while the key management channel (green arrow) organises these secret keys and supplies them upon demand to different encryption applications. QKD: Quantum key distribution.}
    \label{fig:basic QKD}
\end{figure}

The basic functionalities of a general QKD commercial system are illustrated in \autoref{fig:basic QKD}. The QKD system is made up of two parts: a QKD transmitter and a QKD receiver, commonly known as Alice and Bob, respectively. Alice and Bob will generate identical random bit strings as QKD-keys based on the underlying protocol consisting of two stages: raw data exchanges over the quantum channel and post-processing over the classical channel to produce symmetric secret keys \cite{RevModPhys.81.1301}. Finally, the key management organises these symmetric secret keys for use in different encryption applications via the designated interfaces. The key management channel is required when expanding the QKD network with multiple users, as it can securely distribute and relay these secret keys within this network. 

\begin{figure}[!htbp]
    \centering
    \includegraphics[width=0.9\textwidth]{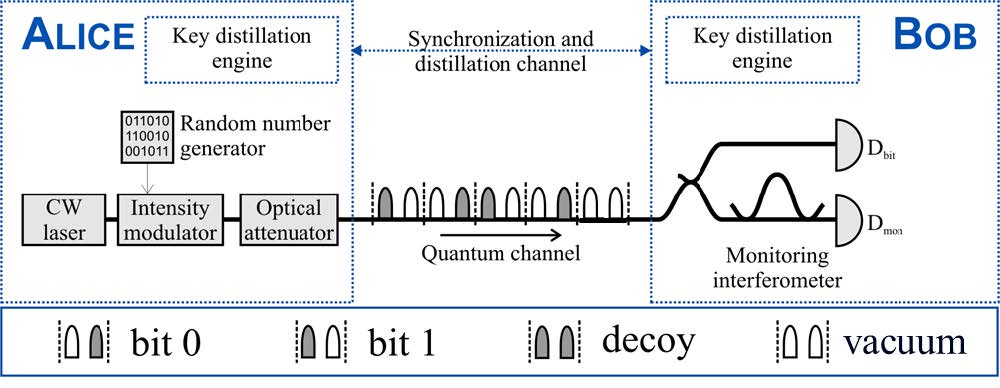}
    \caption{The schematic description of COW-4 protocol. Single-photon-level  pulses with bit value of zeros, ones, decoy and vacuum states are sent from Alice to Bob via the quantum channel. At Bob, the bit-generation detector D$_{\text{bit}}$ generates the QKD keys, while the monitoring interferometer measures the coherence between adjacent pulses at detector D$_{\text{mon}}$. The key distillation process commences thereafter and the sifting, error correction, privacy amplification and key management process happen over the classical channel (indicated by the blue dashed arrow) to generate the secret keys. COW: Coherent one-way; QKD: quantum key distribution.}
    \label{fig:COW protocol}
\end{figure}

For this field trial, the QKD equipment vendor collaborating for this demonstration is ID Quantique (IDQ), and the QKD system used is the Cerberis XGR Series \cite{XGR_brochure}. This system has a repetition rate of \SI{1.25}{GHz} \cite{XGR_technical_info} and uses the Coherent One Way (COW) protocol \cite{gisin2004towards, COWprotocol1}, which is patented by IDQ. The implemented COW protocol is secure against restricted types of collective attack~\cite{Branciard_2008,korzh2015provably, walenta2014fast}. The schematic description of the COW protocol is illustrated in \autoref{fig:COW protocol}. In this demonstration, the latest version of the COW protocol with an additional vacuum state, specifically referred to as the COW-4 protocol here, is employed to foil the zero-error attacks against COW protocol \cite{Foilingzero,gao2022simple}. We note that the QKD system deployed is an implementation of a QKD protocol, where there is necessarily a gap with the ideal theoretical QKD protocol (which is ITS) and its realistic implementation.

In the setup, the laser in the transmitter in Alice emits a continuous wave (CW) beam, which is subsequently modulated at the intensity modulator, to provide coherent optical pulses with bit patterns corresponding to the bit value of zeros, ones, decoy and vacuum states. These pulses are attenuated at the optical attenuator to reach single photon levels and travel from Alice to Bob via the quantum channel. In the receiver at Bob, some pulses reach the bit-generation detector, denoted by D$_{\text{bit}}$, through the beam splitter and they are used for generating the QKD keys in the key distillation process. The other pulses, reflected by the beam splitter, enter the path containing the monitoring interferometer to measure the coherence between adjacent pulses at the monitoring detector, denoted by D$_{\text{mon}}$, to monitor for the presence of eavesdroppers \cite{XGR_brochure}. For this above process to work, Alice and Bob need to be synchronised through the classical channel. After the pulses are exchanged between Alice and Bob, the key distillation process commences on the classical channel, where the QKD keys are generated through processes such as sifting, error correction with Low Density Parity Code (LDPC) algorithm, and privacy amplification using the Wegman-Carter Strongly Universal Hashing to obtain secret keys that are uniformly random, identical and secure against an eavesdropper \cite{XGR_brochure}. These keys are then forwarded to the key management, where a portion is employed for authenticating the classical channel, while the remainder becomes the secret keys shared among Alice and Bob \cite{constantin2017fpga,yuan201810}.

\subsection{2.2 QKD Deployment}\label{sec: QKD Deployment}
The architecture for the physical setup of the field trial demonstration is depicted in \autoref{fig:QKD architecture}. The quantum, service and key management system (KMS) channels of the two QKD devices are made via optical fibres for transmitting signals with different wavelengths at C-band. The connections for the quantum channel are Subscriber Connector (SC)/Ultra Physical Contact (UPC), whereas each of the service and KMS channel connections are established with Lucent Connector (LC)/UPC (duplex) connecting to a transceiver. All three channels are transmitted through their own dedicated fibres. At each site, the QKD device is set up as follows:  

\begin{figure}
    \centering
    \includegraphics[width=0.9\textwidth]{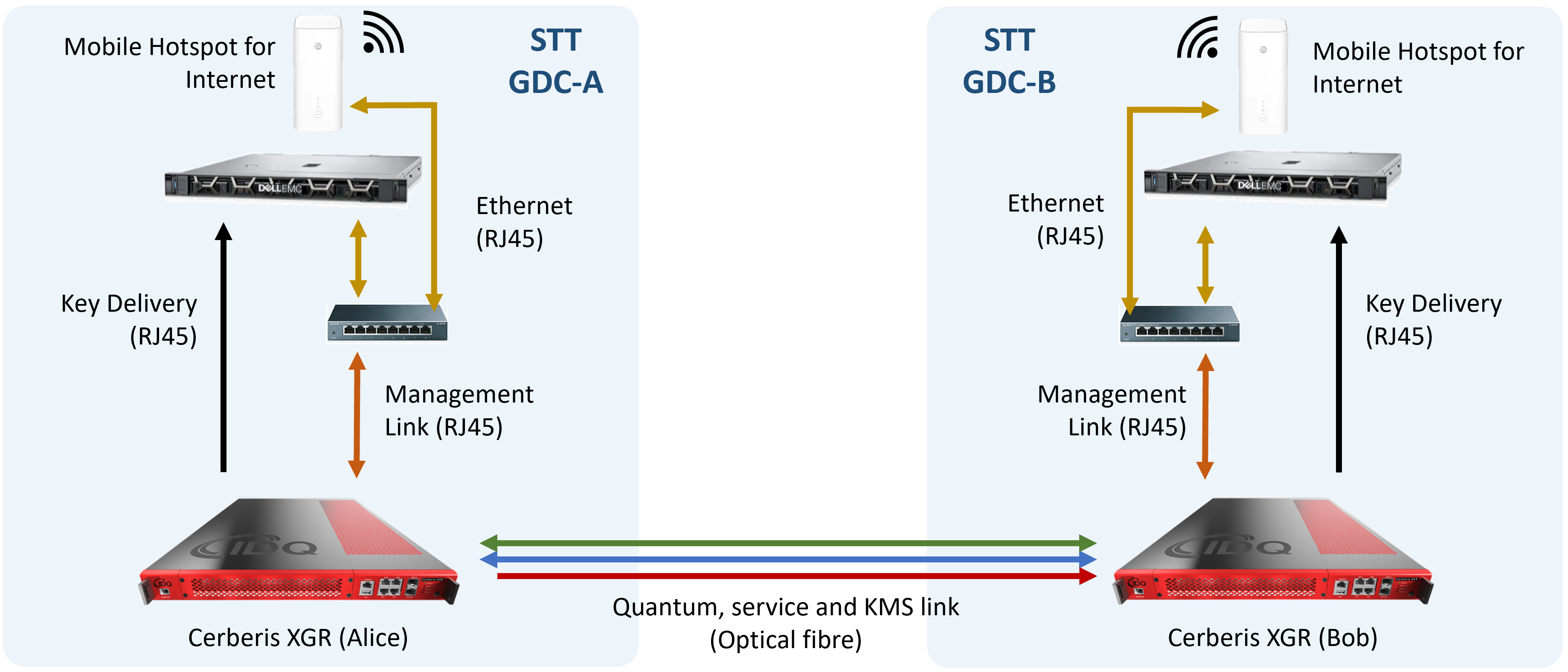}
    \caption{The quantum (red arrow), service (blue arrow) and KMS (green arrow) channels of the QKD systems are connected via optical fibres. The connections established between each QKD system and the IT equipment are made by RJ45 (brown arrow). Within the server, various software containers are deployed, including the QMS, as well as the Web API container, which is managed by the management link (orange arrows). The QKD keys are pushed out of the QKD devices into the server using the ETSI GS QKD 014 REST API via the key delivery link (black arrow). KMS: Key management system; QKD: quantum key distribution; IT: information technology; QMS: Quantum Management System; API: application programming interface; STT GDC: Singapore Technologies Telemedia Global Data Centres.}
    \label{fig:QKD architecture}
\end{figure}

\begin{enumerate}
    \item Two RJ45 Ethernet connections are available from the IDQ Cerberis XGR device, one for the key delivery link and the other for the management of the appliances. The connection for the appliance management is extended into an Ethernet network switch.
    \item  A typical server with dual Ethernet connections is subsequently connected to the switch to control the management of the appliances, while the other is used to control the key delivery link.
    \item Within the server, various software containers are deployed. Important containers include the quantum management system (QMS), as well as the Web API container \cite{XGR_brochure}, which can set up, control, and monitor the QKD devices, and the ETSI GS QKD 014 \cite{ETSI-014} Representation State Transfer (REST) API container that pushes the secret keys out of the QKD devices via the key delivery link. 
    \item A mobile hotspot router is used for internet purposes. The router utilises publicly available consumer mobile networks with either 4G or 5G connectivity. The router is connected to the switch for the management network. We note that the cellular networks are from different operators, and there is no direct communication link setup between both locations, hence, both locations reside in two separate IP networks. 
\end{enumerate}

In our demonstration, the QKD deployment employs the trusted node configuration. In particular, the server and the QKD, which is connected via the ETSI GS QKD 014 REST API interface, are co-located within a trusted environment. We note that in this trusted node setting, we have assumed that the eavesdroppers do not have access to the ETSI GS QKD 014 API interface. The performance of the deployed QKD over time is evaluated and the most critical parameters to monitor are the following:

\begin{itemize}
    \item \textbf{QBER.} This is defined as the ratio of non-identical bits between the QKD transmitter (Alice) and QKD receiver (Bob), which is an error rate due to the quantum signal transmitting via the open quantum channel. In the security proof, all errors are attributed to the eavesdropping action on the open channel. In other words, QBER directly impacts the final SKR, and thus the QKD channel security and performance. A high QBER will result in the aborting of QKD protocol \cite{djordjevic2019physical}, and for a given QKD protocol, a certain threshold of QBER forbids the QKD protocol to generate any secret key. For instance, in a BB84 QKD protocol, QBER$\approx 11\%$ is the theoretical limit to have a positive key rate \cite{christandl2004generic,Shor_2000}. 
    In practice, the noise presence in the quantum channel and imperfection of realistic QKD transmitter and receiver will also contribute to QBER. 
    \\
    
    \item \textbf{SKR.} This indicates the amount of secret keys that can be generated per time period, with the unit of bits per second (bps). Here, the SKR for the COW protocol is derived as a function of QBER, other security parameters and accounting for various eavesdropping attack models, such as sequential \cite{PhysRevLett.125.260510}, collective \cite{Branciard_2008} or zero-error attack \cite{Trényi_2021} that could potentially be executed on the QKD system \cite{RevModPhys.81.1301,djordjevic2019physical}. In practical implementations, the actual SKR will involve a real-time QKD operation time period, which includes the time for the key distillation and post-processing\cite{constantin2017fpga,yuan201810}. Generally, a QKD system with a higher SKR will have a greater advantage in supporting applications that consume keys rapidly.
\end{itemize} 

\subsection{2.3 Fibre Network Infrastructure}
\begin{figure}[!htbp]
    \centering
    \includegraphics[width=0.8\textwidth]{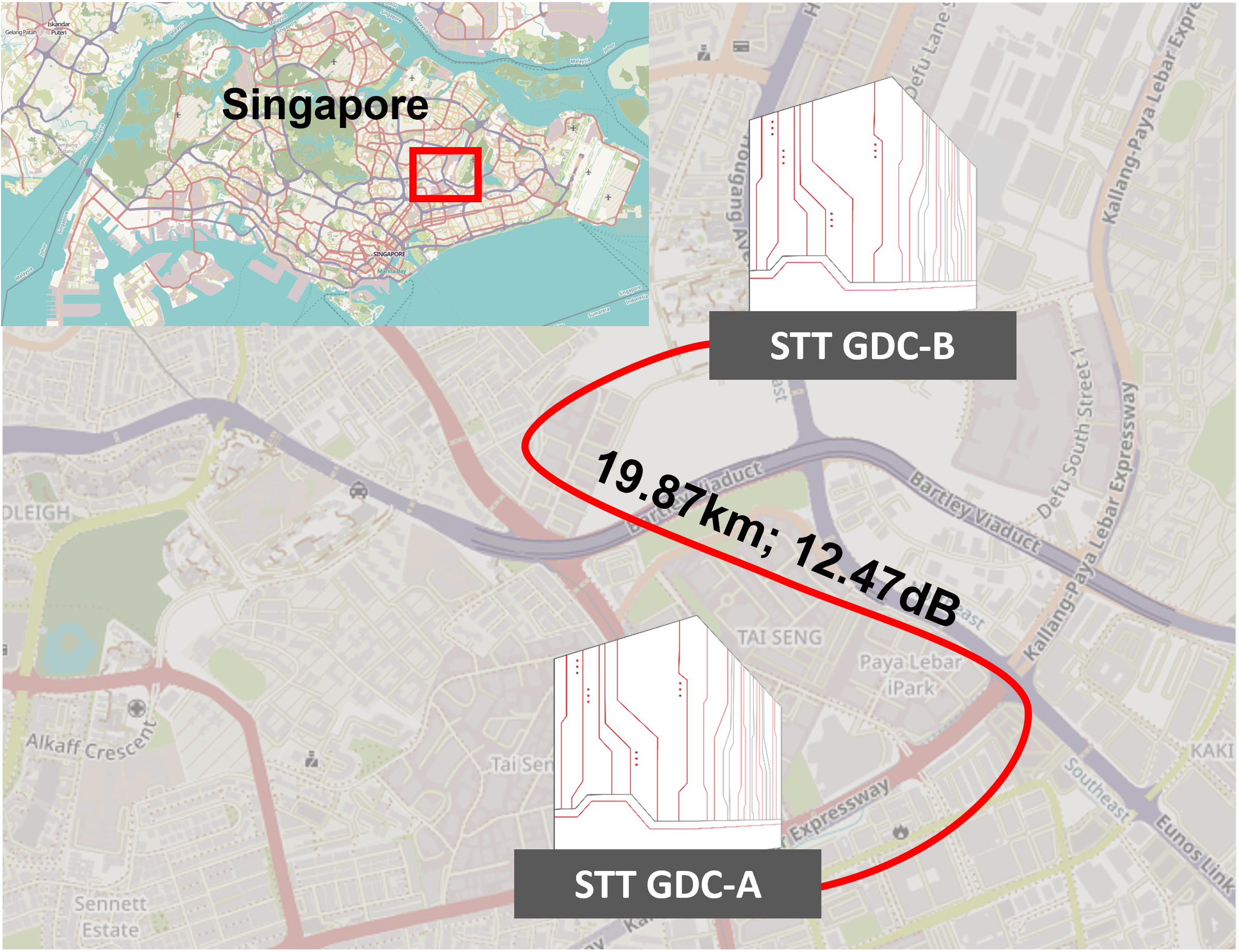}
    \caption{Location of QKD devices at STT GDC-A and STT GDC-B shown on the map. The red line indicates the fibre connection between the two sites with a measured length of \SI{19.87}{km} and a measured loss of \SI{12.47}{dB}. QKD: Quantum key distribution; STT GDC: Singapore Technologies Telemedia Global Data Centres.} 
    \label{fig:QKD map}
\end{figure}

The production-grade fibre network infrastructure is provided by NetLink Trust (NLT). As mentioned in the Introduction section, minimal alterations need to be made to this existing fibre network in this deployment. In particular, apart from the establishment of the last-mile connectivity, the QKD devices can be connected to the existing fibre networks without laying new fibre cables in between. Based on theoretical fibre study conducted by NLT to estimate the fibre length distance and fibre loss via the network planning tool, two data centres from STT-GDC are chosen for this trial demonstration, where QKD Alice and Bob are located at STT GDC-A and STT GDC-B, respectively. The location of the two sites is depicted in \autoref{fig:QKD map}. The fibre connection consists of a total of seven hops from STT GDC-A to STT GDC-B that are in compliance with G.657A standards \cite{G.657A} and the G.652D standards \cite{G.652D}. Upon completion of the fibre connection, an end-to-end optical time-domain reflectometer (OTDR) measurement on the final fibre cable link is conducted. The equipment used in the measurement is a VIAVI SmartOTDR measurement tool. The measured fibre cable link results between these two data centres are shown in \autoref{tab:Fibre Study}. These results are crucial to ensure the QKD equipment continues to operate within its acceptable capability and operation range. 

\begin{table}[!h]
  \centering
  \caption{Results for fibre network infrastructure for the optical fibres between the two data centres}
  \label{tab:Fibre Study}
  \begin{tabular}{ccccccc}
    \toprule
    \begin{tabular}[c]{@{}c@{}}Location 1\end{tabular} & \begin{tabular}[c]{@{}c@{}}Location 2\end{tabular}  & \begin{tabular}[c]{@{}c@{}}Measured Fibre Length \\(OTDR) \end{tabular} & \begin{tabular}[c]{@{}c@{}}Measured Fibre Loss at \SI{1550}{nm}\\(OTDR)\end{tabular} \\
    \midrule
    STT GDC-A & STT GDC-B &  \SI{19.87}{km} & \SI{12.47}{dB} \\
    \bottomrule
    \multicolumn{5}{l}{\begin{tabular}[l]{@{}l@{}}OTDR: Optical time-domain reflectometer; STT GDC: Singapore Technologies Telemedia Global\\ Data Centres.
    \end{tabular}}
  \end{tabular}
\end{table}

\subsection{2.4 QKD Application}

To emulate a secure file transfer using the QKD system, we demonstrate the application of using symmetric QKD-generated keys to encrypt and decrpyt the data that is sent across the two data centres. To this end, a quantum-secured VPN (Q-VPN) application is deployed over the connecting sites. This Q-VPN consumes the QKD keys from the key buffer storage in the QKD system and performs Advanced Encryption Standard-256 (AES-256) encryption thereafter to establish a quantum-safe VPN tunnel for secure data transfer. AES-256 is a symmetric algorithm that is quantum-safe and remains secure even against quantum attacks such as Grover's algorithm. When used in conjuction with a QKD protocol, the overall security of the Q-VPN remains quantum-safe at the protocol level. For the purpose of this demonstration of functionality, cloud resources are utilised to establish the VPN tunnel due to their robustness and ease of implementation in a simplified network setting (In practise, the edge cloud resources can also be located at the trusted nodes, as demonstrated by the NQSN in another trial \cite{AWScloud}). 

\begin{figure}[!htbp]
    \centering
    \includegraphics[width=0.8\textwidth]{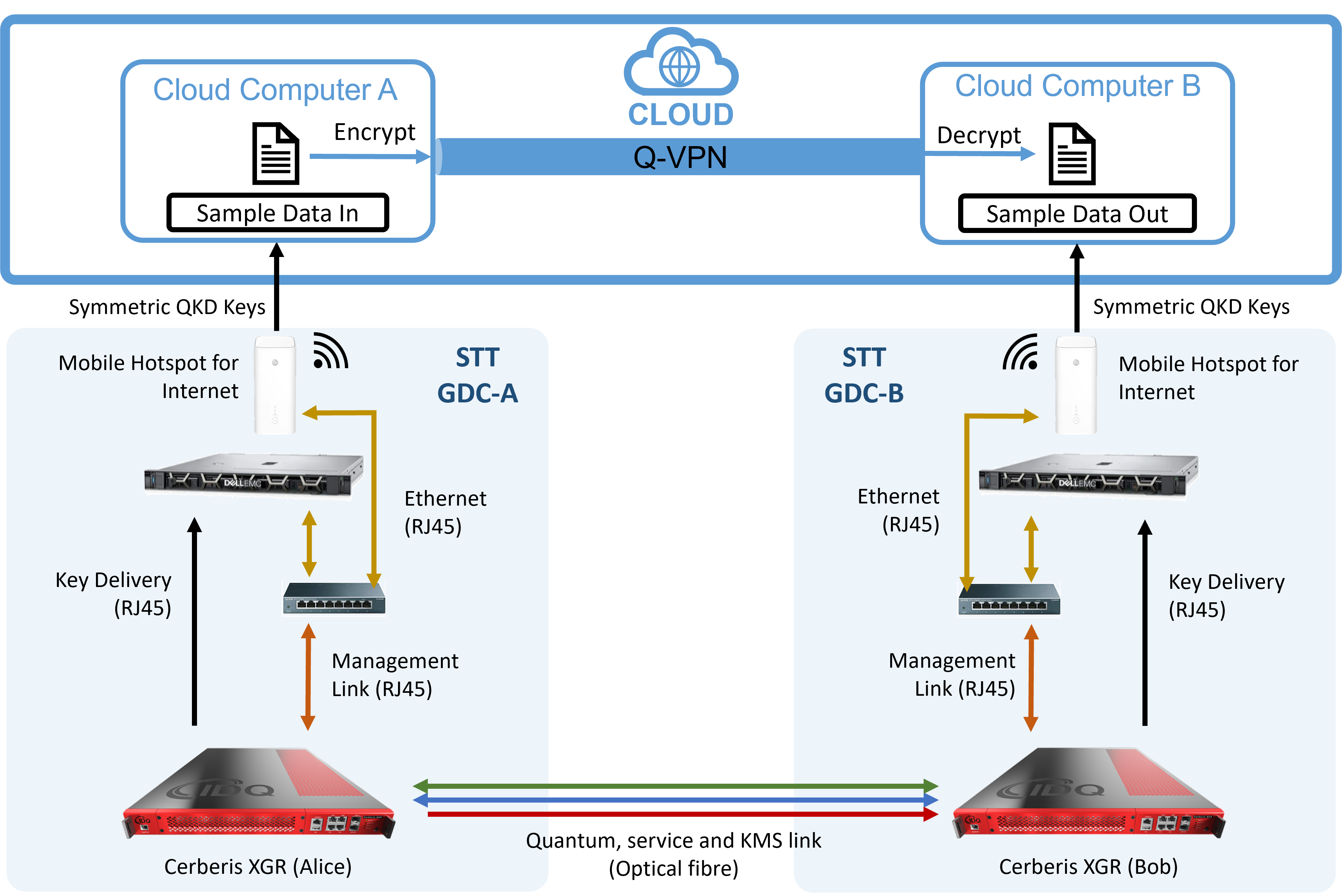}
    \caption{QKD application architecture. The secret keys generated by Alice and Bob are sent to their respective cloud computer via the ETSI GS QKD 014 REST API with the help of the key delivery link (black arrow) and the Ethernet link (brown arrow). These secret keys are stored in the internal SQLite3 database of the Q-VPN application. These stored keys are synced between the two points to encrypt the link between them. Once the Q-VPN link is established, a file transfer client/server application can be achieved.}
    \label{fig:Application layer}
\end{figure}

The architecture of the QKD application is illustrated in \autoref{fig:Application layer}. The symmetric secret keys generated by both Alice and Bob are channelled via the ETSI GS QKD 014 REST API to the server, then onto the cloud computers.  After the Transport Layer Security (TLS) handshake is established between Alice and Bob, these secret keys are stored in the internal SQLite3 database of the Q-VPN application. We note that the connection between the server and the cloud is not assumed to be quantum-safe and only for demonstration purposes. These stored keys are synced between the two points and can be accessed to encrypt the network between them using AES-256 encryption, with the keys renewed every 10 seconds. The used keys will be discarded and the Q-VPN will request more secret keys from the QKD devices to replenish keys in its database. A file transfer client/server application can be achieved. The data is transferred from the sender to the receiver using the Secure Copy Protocol command (SCP) between the cloud computers via this encrypted Q-VPN tunnel.

\section{3. Results}
\subsection{3.1 Performance and reliability analysis of the QKD equipment}

\begin{figure}[!htbp]
    \centering
    \begin{subfigure}{0.49\textwidth}
        \centering
        \includegraphics[width=\textwidth]{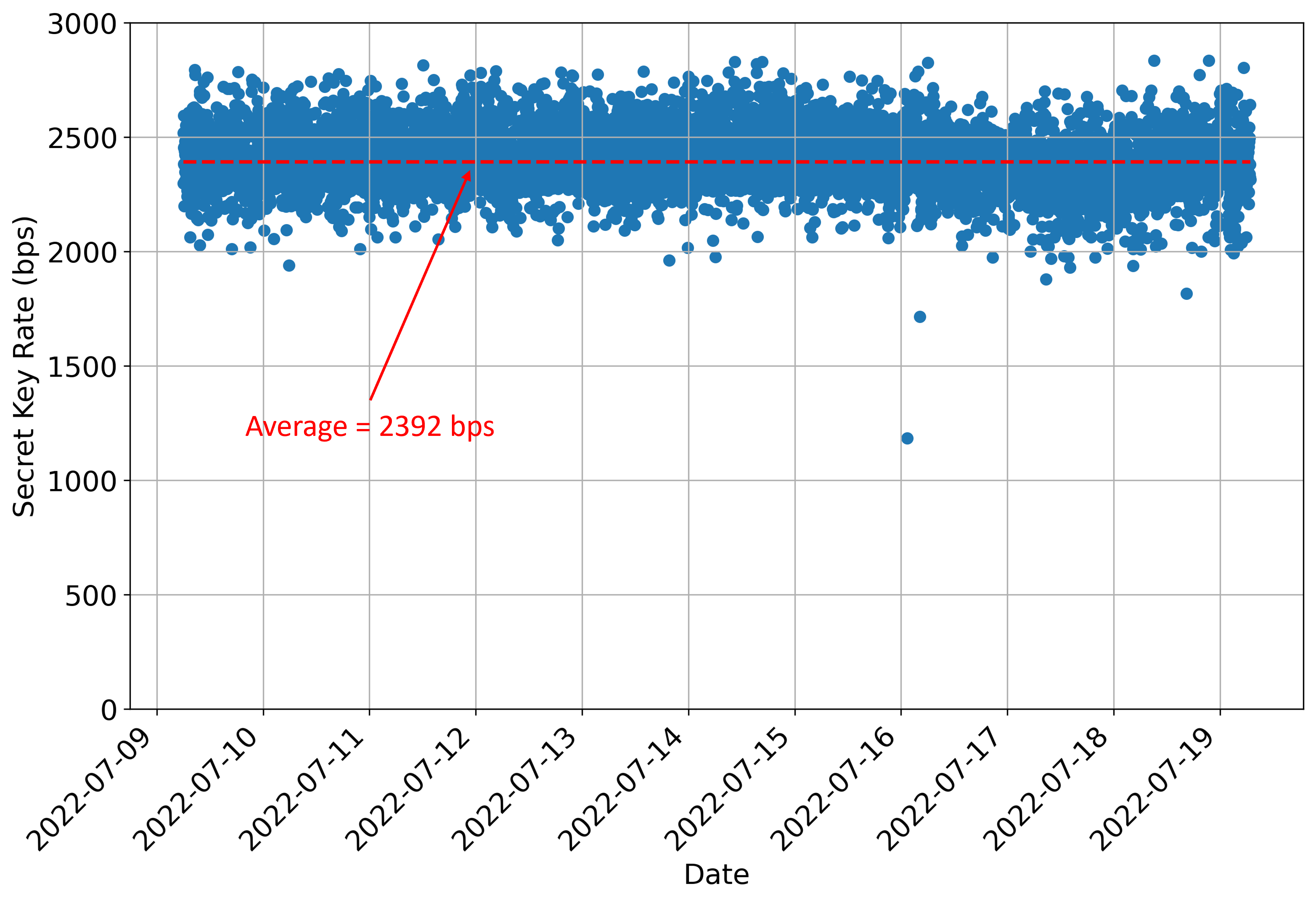}
    \end{subfigure}
    \centering
    \begin{subfigure}{0.49\textwidth}
        \centering
        \includegraphics[width=\textwidth]{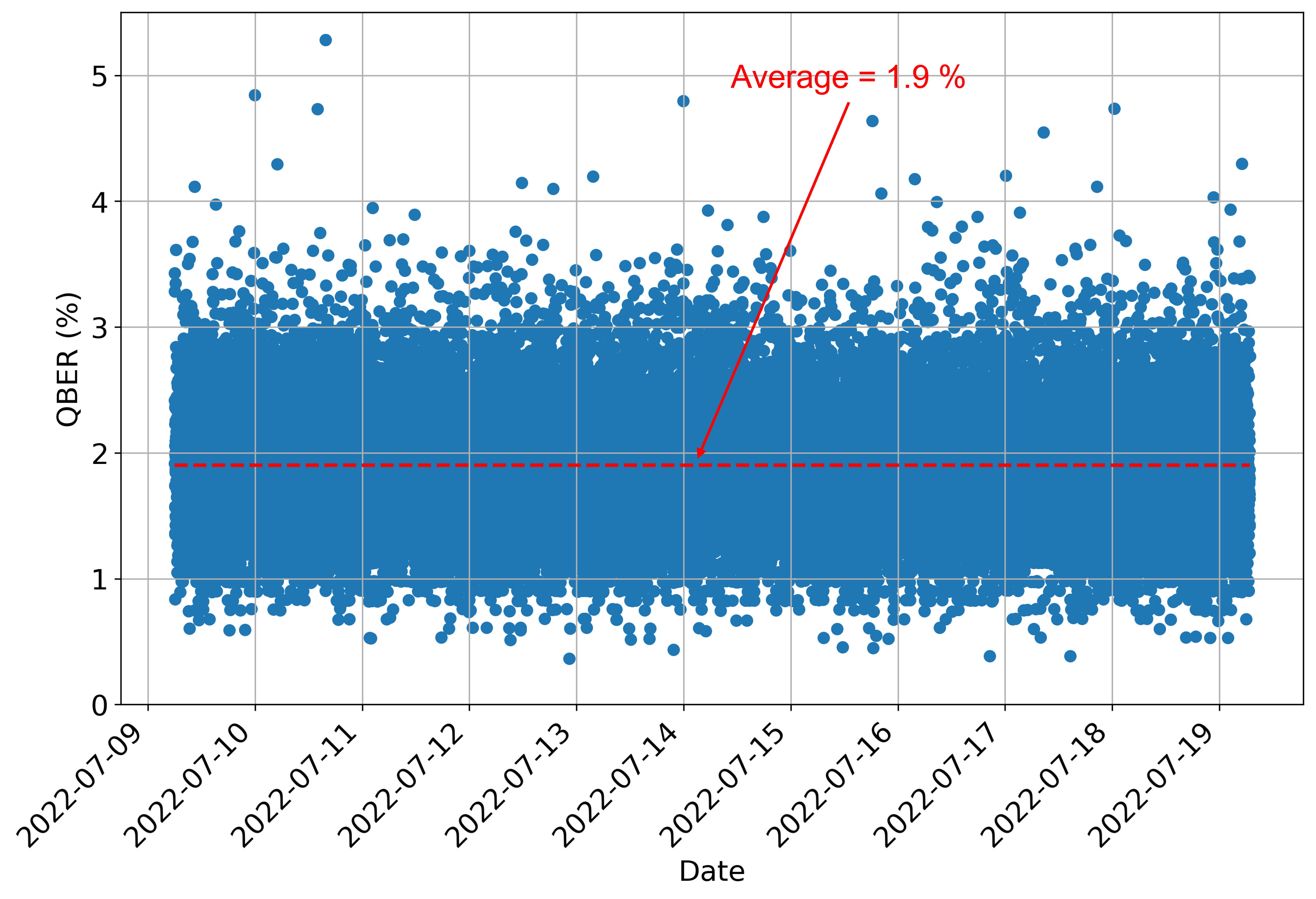}
    \end{subfigure}
    \centering
    \begin{subfigure}{0.49\textwidth}
        \centering
        \includegraphics[width=\textwidth]{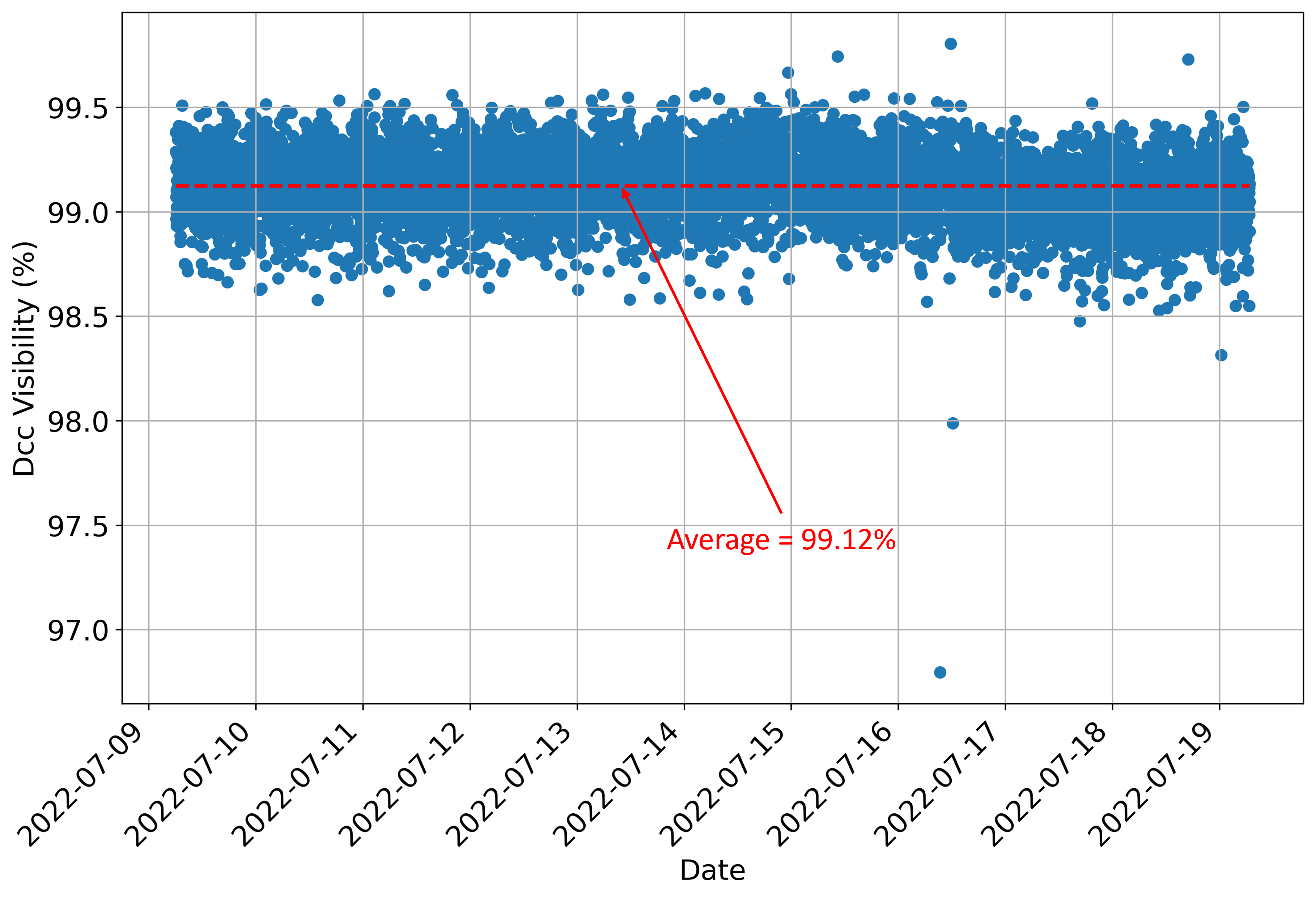}
    \end{subfigure}
    \centering
    \begin{subfigure}{0.49\textwidth}
        \centering
        \includegraphics[width=\textwidth]{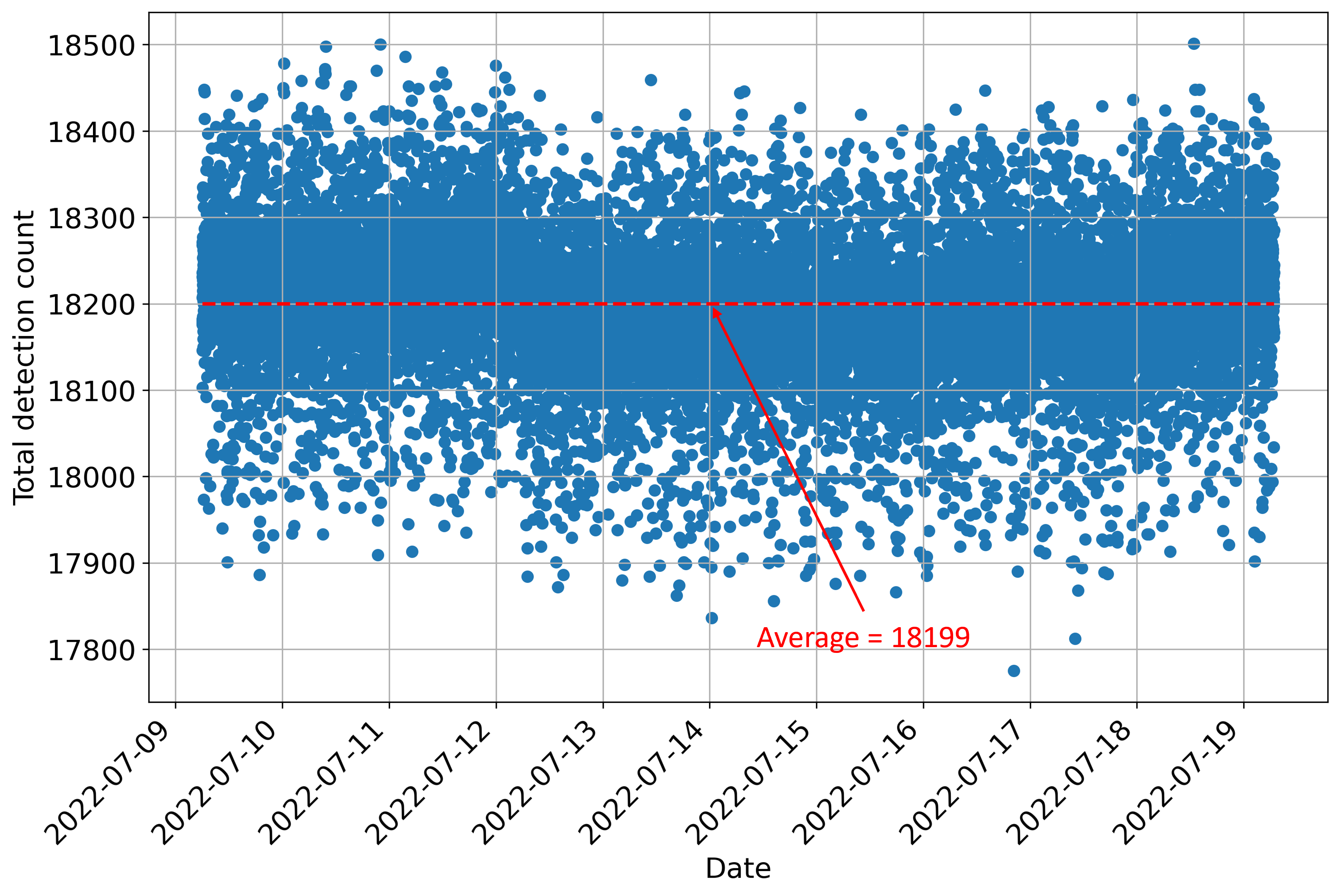}
    \end{subfigure}
    \caption{The SKR, QBER, Visibility and Total Detection Count measurement over a period of 10 days. (Top left) The recorded SKR over time, with an average SKR of \SI{2392}{bps} (indicated by the red dotted line); (Top right) The recorded QBER over time, with an average QBER of \SI{1.9}{\%}(indicated by a red dotted line) and all the recorded QBER is less than \SI{6}{\%}; (Botton left) The Dcc Visibility recorded over time, with an average of \SI{99.12}{\%} (indicated by a red dotted line); (Botton right) The Total Detection Count recorded over time, with an average of \SI{18199}{} (indicated by a red dotted line). SKR: Secret key rate; QBER: quantum bit error rate; Dcc: dark-count corrected.}
    \label{fig:SKR and QBER}
\end{figure}

The stability of SKR and QBER is monitored continuously, and the results are shown in \autoref{fig:SKR and QBER}. The SKR and QBER from the QKD equipment are relatively consistent throughout the time window of ten days. The standard deviation of SKR is less than \SI{4.3}{\%} of its average, indicating a relatively stable performance over the operating period. Meanwhile, the average QBER is relatively low at \SI{1.9}{\%}. More than \SI{97}{\%} of the data points recorded are below \SI{2.9}{\%} of QBER, and maximum recorded QBER is less than \SI{6}{\%}. The total key generated amounts to more than 2 Gigabits, or equivalently more than 8 million AES-256 keys. 

There are two other parameters that are crucial to the performance and reliability specifically to the COW protocol: the dark-count corrected (Dcc) Visibility and the Total Detection Count. The Dcc Visibility is the interference visibility detected at the $\text{D}_{\text{mon}}$ after correcting for the dark counts (detection without incident light). Preferably, the Dcc Visibility should be near, but strictly below, \SI{100}{\%} and the recorded result is sufficiently high, averaging $\SI{99.12}{\%}$ with a small standard deviation of $\SI{0.16}{\%}$. The Total Detection Count accounts for the total number of detection, including incoming photons and dark counts at $\text{D}_{\text{bit}}$. The respective plots for the Dcc Visibility and Total Detection Count are also presented in \autoref{fig:SKR and QBER}. The average and the standard deviation of these parameters are presented in the \autoref{tab:Average SKR and QBER}.

\begin{table}[!htbp]
  \centering
  \caption{Average and standard deviation of the key parameters for the stability test}
  \label{tab:Average SKR and QBER}
  \begin{tabular}{ccc}
    \toprule
    \begin{tabular}[c]{@{}c@{}}Key Parameters\end{tabular} & \begin{tabular}[c]{@{}c@{}}Average\end{tabular} & \begin{tabular}[c]{@{}c@{}}Standard Deviation\end{tabular} \\
    \midrule
    Secret Key Rate & \SI{2392}{bps} & \SI{126}{bps} \\
    QBER & \SI{1.90}{\%}  & \SI{0.50}{\%} \\
    Dcc Visibility & \SI{99.12}{\%}  & \SI{0.16}{\%}\\
    Total Detection Count & \SI{18199}{}  & \SI{65}{}\\
    \bottomrule
    \multicolumn{3}{l}{\begin{tabular}[l]{@{}l@{}}QBER: Quantum bit error rate; Dcc: dark-count corrected.
    \end{tabular}}
    \end{tabular}
\end{table}

\subsection{3.2 Attenuation test on key parameters}

\begin{figure}[!htbp]
  \centering
  \includegraphics[width=0.45\textwidth]{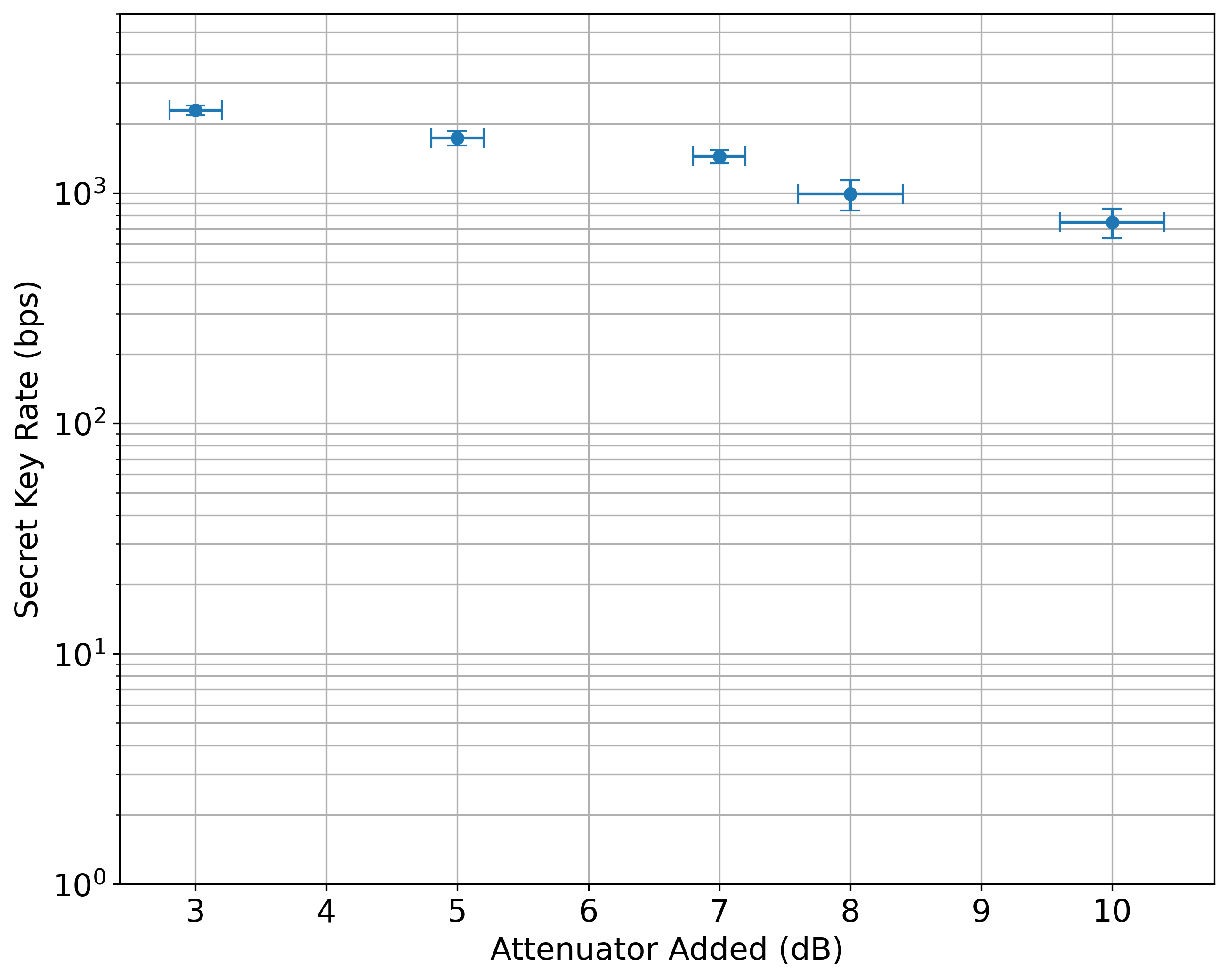}\hspace{1mm}
  \includegraphics[width=0.45\textwidth]{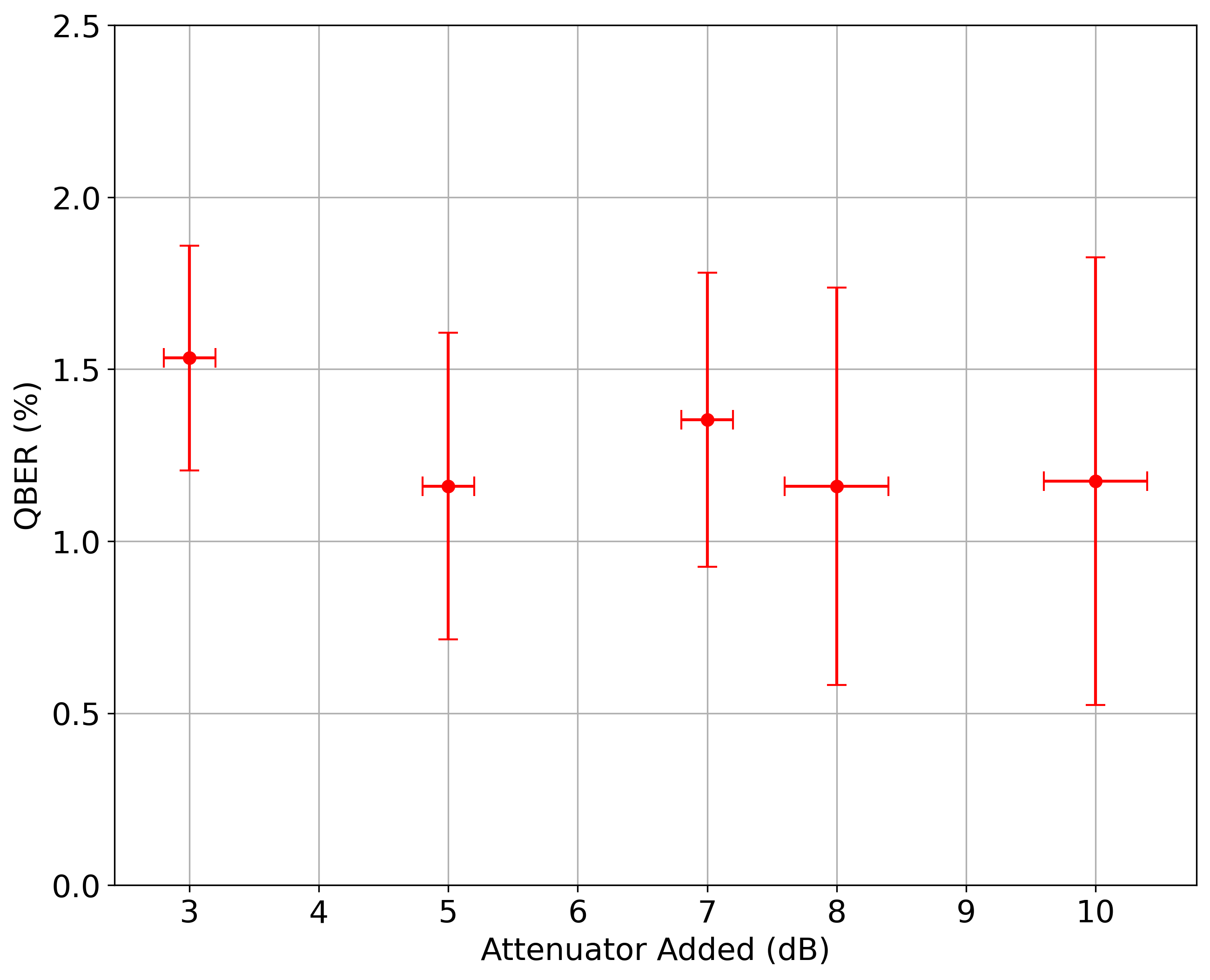}\\[0.7mm]
  \caption{The graphs of attenuation added and its impact on the respective key parameters. (Left) SKR against Attenuator(s) Added. The SKR axis scales in a logarithmic manner to illustrate the relationship between the SKR and added attenuation. The standard deviation for the SKR is included in the figure as error bars. The error bars are relatively small, indicating a relatively constant SKR; (Right) QBER against Attenuator Added. The standard deviation for the QBER is included in the figure as error bars. In both plots, the error bars for the attenuation account for the insertion loss uncertainties, with the larger uncertainty indicating a combination of two attenuators. SKR: Secret key rate; QBER: quantum bit error rate.}
  \label{fig:Attenuator plot}
\end{figure}

An attenuation test is done to ascertain the relationship between the attenuator added and the two key parameters. This analysis could provide an estimation of the potential QKD performance on the fibre for scenarios of different distances. The attenuation is added using fixed attenuators to the optical fibre for the quantum channel. \autoref{fig:Attenuator plot} illustrates the combined results obtained for the key parameters and \autoref{tab:Attenuator added} shows the values of the key parameters with the respective attenuation added. The SKR decreases with the increase in attenuation. During the test, we have further added an attenuation of \SI{12}{dB} , which has resulted in a zero key rate. This indicates that the loss value has exceeded the QKD system tolerable limit.

\begin{table}
  \centering
  \caption{Attenuation test for key parameters}
  \label{tab:Attenuator added}
  \begin{tabular}{ccc}
    \toprule
    \begin{tabular}[c]{@{}c@{}}Attenuator added (dB)\end{tabular} & \begin{tabular}[c]{@{}c@{}}Secret key rate (bps)\end{tabular} & \begin{tabular}[c]{@{}c@{}}QBER (\%)\end{tabular} \\
    \midrule
    \SI{3}{} $\pm$ \SI{0.2}{} & \SI{2303}{} $\pm$ \SI{135}{} & \SI{1.62}{} $\pm$ \SI{0.84}{} \\
    \SI{5}{} $\pm$ \SI{0.2}{} & \SI{1730}{}  $\pm$ \SI{126}{} & \SI{1.13}{} $\pm$ \SI{0.48}{} \\
    \SI{7}{} $\pm$ \SI{0.2}{} & \SI{1473}{}  $\pm$ \SI{141}{} & \SI{1.30}{} $\pm$ \SI{0.47}{} \\
    \SI{8}{} $\pm$ \SI{0.4}{} & \SI{1016}{}  $\pm$ \SI{164}{} & \SI{1.13}{} $\pm$ \SI{0.59}{} \\
    \SI{10}{} $\pm$ \SI{0.4}{} & \SI{746}{}  $\pm$ \SI{110}{} & \SI{1.19}{} $\pm$ \SI{0.66}{} \\
    \bottomrule
  \end{tabular}
\end{table}

\subsection{3.3 QKD Application Integration}
The QKD application showcases a secure file transfer via the Q-VPN tunnel from Alice to Bob (without additional attenuation). The sample files are successfully encrypted and transferred through the Q-VPN every minute, and the content is successfully decrypted at the receiving end. Since the Q-VPN uses AES-256 encryption, the average SKR generated by the QKD devices can provide an AES-256 key refresh rate of 11 keys per second. Given that the Q-VPN renews its key every ten seconds, the QKD devices operating in the commercial environment have the capability to generate sufficient keys to support the operation of the Q-VPN tunnel.

\section{4. Discussion \& Outlook}
The successful demonstration of the QKD keys distributed among the two secured sites, together with a simple application of establishing a Q-VPN, paves the way for a quantum-safe connectivity in real-world use cases and further advanced applications. In the case of the QKD systems, this field trial demonstrates the commercial viability of QKD integration with the existing production-grade fibre network in Singapore within a data centre environment. This is an important milestone, without taking for granted that the data centre environment is ideal for QKD devices. For instance, these QKD devices can be co-located with other telecommunication equipment, including encryptors, servers, plus computational intensive devices, impacting the surrounding temperature stability. In comparison to other works demonstrating a similar QKD protocol in controlled lab settings \cite{expdemo170,korzh2015provably}, by showcasing stable secret keys exchanged, our work bears a better resemblance to an operational environment and provides insight into understanding the practical challenges for the QKD system.

Another important aspect studied in our demonstration covers the organisation of fibre connectivity for the QKD deployment. For this demonstration, there are seven patches across Alice and Bob, giving a measured fibre distance of \SI{19.87}{km} and \SI{12.47}{dB} of fibre loss. Apart from ensuring that the total fibre loss is within the capability and operation range of the QKD devices, no further optimisation is required. In principle, the insertion loss from the fibre patches can be further minimised via fibre splicing to reduce the number of patches. In our demonstration, while dedicated dark fibres are used for the quantum and the classical channels, in principle one can utilise wavelength-division multiplexing (WDM) technique to conserve the fibre resources. For instance, by having the quantum signal operating at a different optical band (e.g. O-band) with respect to the classical data signals, channel multiplexing over a single fibre core can be performed \cite{XGR_brochure}.

In this demonstration, a point-to-point QKD architecture linking two data centres was employed. It is crucial to recognise that these data centres are part of an interconnected network made up of multiple data centres, which ultimately need to be scaled beyond mere point-to-point connections to guarantee quantum safety across the network. With the multi-layer approach of QKD network architecture\cite{RS_QKDN, ITU_Y.3800}, it allows  scalability and interoperability from point-to-point QKD to multi-point QKD network topology. Under the trusted relay node-based QKD network, this is enabled by the key management (KM) layer to interconnect the QKD pairs with key supply interfaces as well as key relaying and storage functions inside KM layers. Beyond trusted nodes, there are other quantum technologies under active development to extend and enhance the QKD network, such as measurement-device-assisted QKD (measurement-device-independent QKD \cite {PhysRevLett.108.130503}, twin-field QKD \cite{lucamarini2018overcoming}) and quantum repeater \cite{RevModPhys.95.045006}. Examples of multi-point QKD topology include a mesh \cite{peev2009secoqc, Wang:14,Sasaki:11}, a ring \cite{dynes2019cambridge}, a star \cite{chen2021implementation,frohlich2013quantum,Fan-Yuan:22} or a mixed type architecture \cite{chen2021integrated}. Moreover, the performance study of different QKD protocols and vendors in the market can also be done to analyse their performance and commercial viability within the data centre or other mission critical infrastructure environment. Some examples of the QKD protocols are the BB84 system, entanglement-based system, and continuous variable system. The utilisation of emerging technologies could further enhance the performance of QKD devices and their realisation could also be examined in future demonstrations. These advancements include fast single photon detectors \cite{grunenfelder2023fast}, integrated transmitter and receiver \cite{Sax:23}, qubit-based time synchronisation technique \cite{huang2024cost} and digital signal processing \cite{matsuura2021finite, chen2023continuous}. Apart from the Q-VPN applications, different use cases at different Open Systems Interconnection (OSI) layers could also be explored in the future.

On the other hand, it is instructive to mention another quantum-safe cryptography alternative, which is the post-quantum cryptography (PQC) \cite{bernstein2017post}. PQC is a cryptographic algorithm that is believed to be secured and resilient under known quantum algorithm attacks. It finds applications in cryptography such as digital signature, public-key encryption and key establishment. PQC, being primarily software based, suggests that quantum-safe migration and implementation could be cost-effective and scalable. However, to maintain a certain degree of performance, hardware upgrade might be required as well. For QKD and PQC, though both offer quantum-safe solution in the post-quantum era, they still require standardisation and certification. This is to ensure that the respective encryption protocols are implemented properly, preventing potential vulnerability and loopholes in their implementation security before its widespread adoption. Here, we provide a high-level comparison between PQC and QKD in \autoref{tab: comparison between PQC and QKD}. A hybrid framework that captures the strengths of the QKD devices and PQC could be implemented to improve the overall resiliency~\cite{renner2023debate,brauer2024linking}.

\begin{table}[!h]
  \centering
  \caption{Comparison between PQC and QKD}
  \label{tab: comparison between PQC and QKD}
  \begin{tabular}{lll}
    \toprule
    \begin{tabular}[c]{@{}c@{}}\end{tabular} & \begin{tabular}[c]{@{}c@{}}\textbf{PQC}\end{tabular}  & \begin{tabular}[c]{@{}c@{}}\textbf{QKD}\end{tabular} \\
    \midrule
    Implementation & Software and hardware & Hardware \\
    Protocol security & Computational complexity & ITS \\
    Implementation loopholes & Exist & Exist \\
    Application and usage & Public-key encryption and key establishment,  Digital signature & Key establishment \\
    Migration & Software and hardware upgrade & Infrastructure and hardware upgrade \\
    Standardisation and certification & Required & Required \\
    \bottomrule
    \multicolumn{3}{l}{\begin{tabular}[l]{@{}l@{}}PQC: Post quantum cryptography; QKD: quantum key distribution; ITS: information-theoretic security.
    \end{tabular}}
  \end{tabular}
\end{table}

\section{5. Conclusions}
We confirmed the feasibility of operating QKD devices over an existing production-grade fibre network within a commercial data centre environment. In terms of the QKD device's performance, the secret key rate and QBER are stable and consistent over the trial period. In particular, we achieved an average SKR of \SI{2.392}{kbps}, which is largely achievable due to a low average QBER of less than \SI{2}{\%}. A total of more than 2 Gigabits of AES-256 keys are accumulated, with the rates of around 690 sets of keys per minute. The attenuation test verifies the functionality of the QKD equipment over different quantum channel losses between Alice and Bob. For the application, files are successfully transferred between two data centres via the Q-VPN which makes use of the secret key generated by the QKD devices. Our efforts mark the inaugural stride towards the widespread deployment of QKD throughout Singapore, thereby bolstering the infrastructure for practical, quantum-safe communication.

\section{Declarations}
\subsection{Acknowledgments}

 We acknowledge ST Telemedia Global Data Centre for providing secured physical locations to host the QKD trial, Netlink Trust for the provisioning of the fibre network and ID Quantique SA for the loan of the QKD system and providing the image in \autoref{fig:COW protocol}. We thank Yu Cai for helpful discussion. Additionally, we are grateful to the Department of Electrical and Computer Engineering, National University of Singapore, for supporting the logistics of the field trial.

\subsection{Authors’ contributions}

Made substantial contributions to conception and design of the study: Qiu K, Haw JY, Qin H, Kasper M

Provisioning and setting up of the field trial environment: Qiu K, Haw JY, Qin H, Kasper M, Ling A

Performed data acquisition, data analysis and interpretation: Qiu K, Haw JY, Qin H

Drafting the manuscript: Qiu K, Haw JY, Qin H, Ng NHY, Ling A

Discussion of the main idea and scientific contribution: Qiu K, Haw JY, Qin H, Ng NHY, Kasper M, Ling A

\subsection{Availability of data and materials}
Not applicable.

\subsection{Financial support and sponsorship}
We acknowledge funding support from the National Research Foundation, Singapore and A*STAR under its Quantum Engineering Programme (National Quantum-Safe Network, NRF2021-QEP2-04-P01) and start-up grant for Nanyang Assistant Professorship awarded to Ng NHY of Nanyang Technological University, Singapore. 

\subsection{Conflicts of interest}

All authors declared that there are no conflicts of interest.

\subsection{Ethical approval and consent to participate}

Not applicable.

\subsection{Consent for publication}

Not applicable.

\subsection{Copyright}

© The Author(s) 2024.

\bibliographystyle{oae}

\end{document}